\documentclass[preprint,12pt, nonatbib]{elsarticle}

\usepackage{amssymb}
\usepackage{amsmath}
\usepackage{gensymb}
\usepackage{array}
\usepackage{upgreek}
\usepackage[colorlinks=true, urlcolor=blue, linkcolor=blue]{hyperref}
\usepackage{scicite}

\usepackage[utf8]{inputenc} 
\DeclareUnicodeCharacter{03B2}{\ensuremath{\beta}}
\DeclareUnicodeCharacter{223C}{\ensuremath{\sim}}

\usepackage[backend=biber, sorting=none, url=false, doi=true]{biblatex}

\begin{document}

\begin{frontmatter}







\title{Accounting for the length-scale dependence of thermal diffusivity in 3C-SiC measured with transient thermal gratings}






\author[label1,label2]{Keshav Vasudeva}
\author[label3]{Samuel Huberman}
\author[label1]{Angus P. C. Wylie}
\author[label4]{Maxwell Rae}
\author[label1,label2]{Joey Demiane}
\author[label3]{Jamal A. Haibeh}
\author[label2,label4]{Elena Botica-Artalejo}
\author[label2]{Kevin B. Woller}
\author[label1]{Michael P. Short}
\author[label2]{Sara E. Ferry}

\affiliation[label1]{organization={Department of Nuclear Science and Engineering, Massachusetts Institute of Technology},
}

    \affiliation[label2]{organization={Plasma Science and Fusion Center, Massachusetts Institute of Technology},
}

\affiliation[label3]{organization={Department of Chemical Engineering, McGill University}}

\affiliation[label4]{organization={Department of Materials Science and Engineering, Massachusetts Institute of Technology}}

\begin{abstract}
Pump-probe optical methods like transient grating spectroscopy (TGS) enable rapid, nondestructive thermoelastic property measurements. But, in phonon-dominated ceramics, they can underpredict bulk thermal diffusivity when long mean free path (MFP) phonons do not equilibrate over experimental length scales. We combine \textit{\textit{\textit{in situ}}} TGS with Si$^{4+}$ ion irradiation of CVD 3C-SiC (300 and 550\degree C, 0.5-1 dpa) and density functional theory informed Boltzmann transport equation solutions to understand the origins of this offset. We show how the discrepancy between laser flash analysis (LFA) and TGS-measured thermal diffusivity varies with grain-boundary density, temperature, and defect concentration. We introduce a dimensionless suppression factor that accounts for this discrepancy and demonstrate its utility by using it to show an agreement between the thermal defect resistance of neutron irradiated 3C-SiC (measured using LFA) and ion irradiated 3C-SiC (measured using TGS). This theory-informed experimental framework enables quantitative, \textit{in situ} tracking of ion irradiation damage induced thermal transport degradation in ceramics. 


\end{abstract}




\end{frontmatter}



\section{Introduction}

The past decade has seen renewed interest in the nuclear industry, with increased investments in both the fission and fusion sectors \cite{FIA_GlobalFusionIndustry_2025, IEA_NewEraNuclear_2025}. The proposed reactors will push materials towards higher-temperature and dose operations than ever before. As a result, the nuclear materials community is actively developing solutions that span both metallic and ceramic-based material candidates \cite{zinkle_materials_2013, zinkle_motivation_2016, kurtz_chapter_2019, sorbom_arc_2015}. 

Ceramics and ceramic-based composites are central to emerging fission designs, with applications ranging from advanced fuels (Tristructural-isotropic or TRISO) to accident-tolerant fuel cladding (SiC/SiC cladding) and in-core sensors (fiber-optic temperature sensors) \cite{lee_opportunities_2013, powers_review_2010, terrani_accident_2018, kandlakunta_silicon_2020, INL_ASI_FY24_Summary_2025}. In addition, SiC and SiC-based composites are candidates for applications in fusion power plants (FPPs) such as structural materials\footnote{The Electric Power Research Institute (EPRI) Fusion Materials Roadmapping Workshop draft report mentions SiC/SiC composite materials as candidate structural materials for FPPs. The draft report is public and has been released for comments. It can be found by following this link: \url{https://www.epri.com/research/programs/065093/events/0b97781d-e3eb-470a-ab11-97800fb84638}.}, fuel channel inserts and tritium permeation barriers \cite{freidberg_liquid_2024, wong_overview_2010, wright_silicon_2015}. Interest in these materials stems from their high temperature stability, high strength, low electrical conductivity, small diffusivities of gaseous products and potential to serve as low-activation substitutes to traditional structural materials 
\cite{katoh_stability_2011, katoh_radiation_2012, snead_limits_2004, snead_handbook_2007, causey_416_2012}.

Given the crystalline structure of these materials, their phonon-dominated thermal diffusivity ($\alpha$) suffers a large drop due to the microstructural defects produced by neutron irradiation \cite{li_atomistic_1998, snead_thermal_2005}. 
As newer materials and manufacturing processes are proposed to address this problem, a complementary experimental approach is needed that rapidly addresses their performance. Ion irradiation offers a potential pathway, permitting damage rates on the order of tens to hundreds of displacements-per-atom (dpa) per day \cite{taller_emulation_2019}. However, the shallow implantation depth of the ions precludes the use of standard, post-irradiation characterization techniques such as laser flash analysis (LFA) and tensile tests. Transient grating spectroscopy (TGS), a laser-based, short-wavelength ($<10 \: \mathrm{\upmu m}$) pump-probe technique, which measures the near-surface thermoelastic properties of materials, can be used \textit{\textit{in situ}} with ion irradiation to provide insights into the evolution of ceramic-based materials as a function of dose \cite{dennett_thermal_2018, dennett_real-time_2019, wylie_accelerating_2025}. In addition to TGS, researchers have investigated other techniques such as time-domain thermoreflectance and modulated photothermal IR radiometry to measure changes in the thermal properties of ceramics on length scales relevant to ion irradiation \cite{scott_phonon_2018, cabrero_thermal_2010}. However, while these provide no insights into elastic properties, they also have limitations in applicability due to mandatory thermal transducer coatings and high temperature interference, respectively. A more detailed review on these techniques is provided in the Supplementary Material.

TGS has been demonstrated as an effective tool to monitor changes in the thermoelastic properties of metallic materials during ion irradiation across a wide temperature range \cite{dennett_capturing_2019, dennett_real-time_2019, wylie_accelerating_2025, aurora_facility_2025}. 
However, previous work has shown that although TGS can be used to measure $\alpha$ of materials with phonon-dominated thermal transport, the measured values are often smaller than the bulk when the experimental length scales (such as TGS thermal grating period) are on the order of, or below, a few micrometers \cite{huberman_unifying_2017, dennett_thermal_2018}. This effect stems from the suppression of phonons with mean free paths (MFPs) longer than experimental length scales due to the transition of these phonon modes from diffusive to ballistic \cite{maznev_onset_2011, hofmann_transient_2019}.
Despite this, recent work has claimed that TGS provides accurate thermal diffusivity measurements of ceramics such as SiC \cite{liu_thermal_2025}. In this paper, we will refine this statement by providing the bounds within which it is true, and suggest methods to interpret thermal diffusivity measurements made using TGS on ceramic samples. Furthermore, many previous studies have explored the interaction between the TGS thermal grating length scales and those relevant to lattice thermal conductivity (set by phonon MFP) \cite{maznev_onset_2011, cuffe_reconstructing_2015, dennett_thermal_2018}. However, these studies largely focus on high-purity ceramics. In this paper, we build on this body of work by showing how the length scales relevant to the thermal processes of 3C-SiC change as a function of grain size, vacancy concentration, and temperature. We then discuss how these changes impact TGS measurements of thermal diffusivity. 

High-purity ceramics, used in fundamental studies of radiation damage and semiconductor manufacturing, are often not optically reflective in the visible light spectrum typically employed in TGS (500 nm - 700 nm) \cite{solangi_absorption_1992}. Prior work had addressed this by depositing reflective gold films ($\sim$ 30 nm). However, given the thermal expansion mismatch between gold and 3C-SiC, reflective coatings could delaminate during experimental studies at the high temperatures relevant to fusion and fission \cite{loiacono2022, radue2018hot, snead_handbook_2007, pamato_thermal_2018}. 
We propose an alternative reflective coating that is stable at high temperatures and yields TGS signal traces that are easier to analyze using the existing fitting function \cite{short2025tgs, dennett_capturing_2019, wylie_accelerating_2025}. Following this. we use TGS to capture the degradation of $\alpha$ of CVD 3C-SiC as a function of temperature, and \textit{\textit{in situ}} during Si$^{4+}$ ion irradiation at 300 and 550\degree C. 
Then, using density functional theory (DFT) informed Boltzmann transport equation (BTE) solutions, we quantify how experimental length scales, grain boundaries, temperature and point defects jointly shape the thermal response of a  TGS signal acquired on 3C-SiC, and identify when TGS measured thermal diffusivity is in correspondence with bulk-relevant transport. Grounded in these results, we propose a dimensionless suppression factor and use it in the calculation of the thermal defect resistance of ion irradiated CVD 3C-SiC. We conclude by comparing the resulting data with prior literature for neutron-irradiated CVD 3C-SiC.

These results establish a practical, physics-based workflow for quantitative screening of irradiation-induced thermal transport degradation in ceramics. While this workflow could enable researchers to compress evaluation timelines of ceramics materials from years to days, it also cautions that data acquired from techniques like TGS should be investigated past the fitted values.

\section{Results}

\subsection{TGS signal of 3C-SiC obtained with gold and tungsten reflective coatings}\label{sec:reflective_coating_results}

TGS relies on the ability of a material to absorb the pump laser power in the near-surface region to generate a surface-localized temperature grating. 
High purity 3C-SiC has a wide band gap and correspondingly low absorption, resulting in large optical penetration depths ($\sim \: 10 \: \mathrm{\upmu m}$) for the visible light spectrum ($\sim \: 500 \: \mathrm{nm}$) \cite{noauthor_nsm_nodate}. Consistent with this, no measurable TGS signal was obtained when 532 nm (pump) and 577 nm (probe) beams were incident on uncoated 3C-SiC. Furthermore, even if a signal was visible, the mathematical formulation behind TGS assumes surface energy absorption \cite{kading_transient_1995, dennett_thermal_2018}. This assumption breaks down when the attenuation depth is on the order of several microns or greater. A coating which allowed near-surface absorption (within the visible light spectrum) was therefore required.

\begin{figure}[h]
    \centering
    \includegraphics[width=0.75\linewidth]{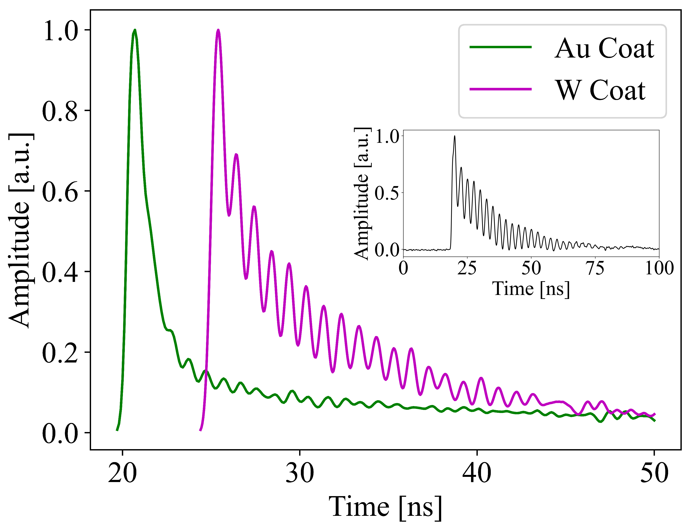}
    \caption{Normalized TGS signals for CVD 3C-SiC coated with approximately 36 nm of Au (sputter coating) and 20 nm of W (PVD). These signals were acquired at room temperature and in vacuum. Note that the temporal shift is only for visual purposes. Inset showing TGS signal trace for single crystal tungsten with {100} surface orientation.}
    \label{fig:Au_vs_W_coat}
\end{figure}

Gold (Au), commonly used as a transducer material in TDTR due to its temperature sensitive reflectivity, was first evaluated \cite{ueji_situ_2020}. However, Au presents practical limitations for phase-grating TGS on SiC. First, its strong thermoreflectance produces a pronounced initial spike in the signal followed by a relatively low-amplitude tail (Fig. \ref{fig:Au_vs_W_coat}). The resulting TGS signal trace obtained with $\sim \: 30 \: \mathrm{nm}$ gold coating on CVD 3C-SiC resembles an amplitude grating signal, making it difficult to isolate and maximize the phase-grating component \cite{dennett_thermal_2018}. The inset in Figure \ref{fig:Au_vs_W_coat} shows a TGS signal trace captured from the tungsten calibration sample (no reflective coating used) where the phase grating response was maximized. It should be noted that the thickness of the gold coating was selected such that it was larger than the expected attenuation depth (calculated based on the dielectric coefficients provided in \cite{werner_optical_2009}), but small enough to minimally interfere with the signal. Refer to the Supplementary Material for this calculation. Second, Au is thermomechanically incompatible with 3C-SiC as their coefficients of thermal expansion differ substantially. This can generate thermal stresses and promote coating delamination at elevated temperatures \cite{snead_handbook_2007, pamato_thermal_2018}.

Tungsten (W) has been widely used in phase-grating TGS measurements \cite{wylie_accelerating_2025, reza_non-contact_2020}, and exhibits a smaller thermal expansion mismatch with SiC than Au. W was therefore tested as an alternative coating \cite{kozyrev_thermodynamic_2023}. In practice, W-coated ($\sim 20 \: \mathrm{nm}$) SiC produced a phase-grating response that was easier to maximize, indicating a reduced thermoreflectance for the W coating as compared to Au. This enabled reliable extraction of both $\alpha$ and SAW speed ($c_R$).  
These results suggest that thin W coatings can enable phase-grating TGS thermoelastic property measurements on otherwise optically transparent or weakly reflective materials. Refer to the Supplementary Material for a discussion on how the thickness of the W coating was selected.

\subsection{Using TGS to capture thermal diffusivity and SAW speed as a function of temperature} \label{sec:thermomechanical_temp}

\begin{figure}[h!]
    \centering
    \includegraphics[width=\linewidth]{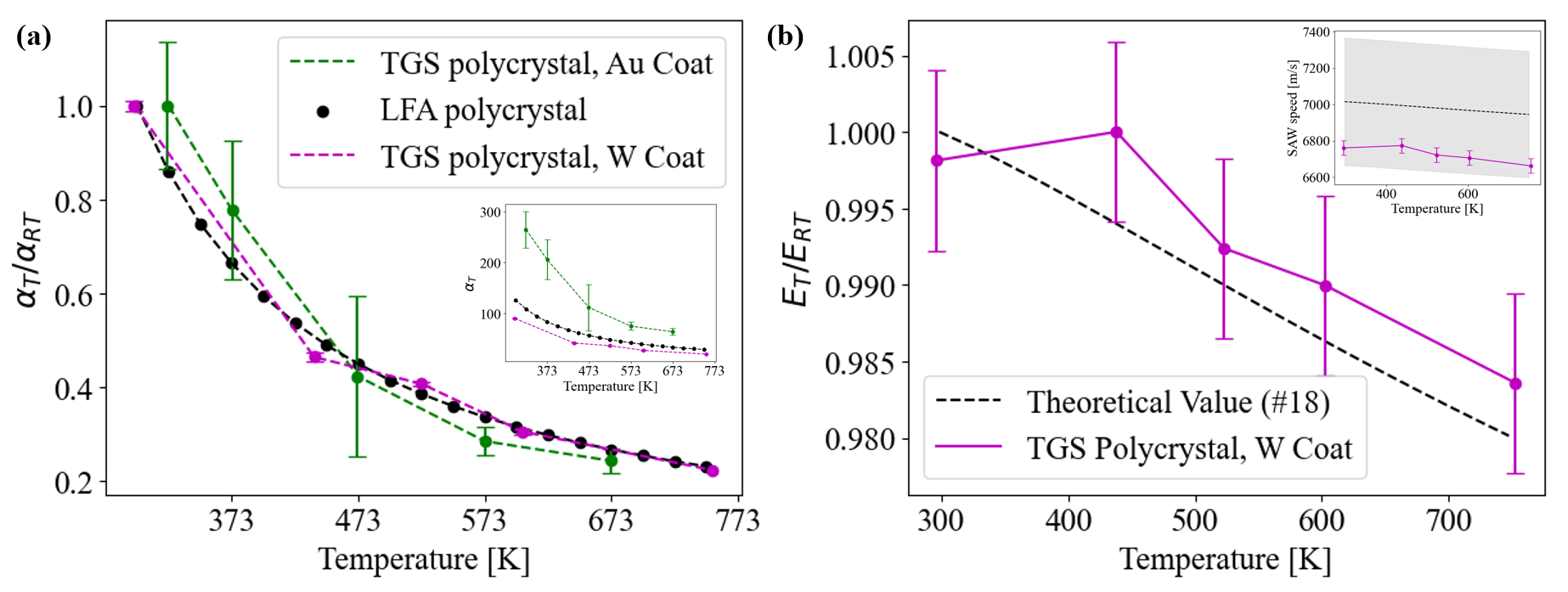}
    \caption{(a) Thermal diffusivity vs temperature, normalized by room temperature thermal diffusivity, measured using TGS (Au- and W-coated SiC) and LFA (SiC). Inset showing absolute values of thermal diffusivity as a function of temperature. (b) Elastic modulus vs temperature, normalized by room temperature elastic modulus, measured using TGS (W-coated SiC) compared to theoretical values (SiC) \cite{snead_handbook_2007}. Inset showing absolute values of SAW speed as a function of temperature. Shaded region in inset shows $\pm \: 5\%$ bounds on the theoretical value}
    \label{fig:temperature_dep_alpha_and_SAW}
\end{figure}

Figure \ref{fig:temperature_dep_alpha_and_SAW} (a) shows that there is good agreement between the $\alpha$ measured using LFA ($\alpha_\mathrm{LFA}$) and TGS ($\alpha_\mathrm{TGS}$) of Au-coated and W-coated SiC when these values are normalized by the room temperature values. Despite the agreement seen between these normalized values, there are some key differences. The inset in Figure \ref{fig:temperature_dep_alpha_and_SAW} (a) shows that $\alpha_\mathrm{TGS}$ of Au-coated SiC is higher than $\alpha_\mathrm{LFA}$ at all temperatures while it is lower for W-coated SiC. Further, $\alpha_\mathrm{TGS}$ of Au-coated SiC has larger error bars than $\alpha_\mathrm{TGS}$ of W-coated SiC and $\alpha_\mathrm{LFA}$. The error bars in Figure 2 denote one standard deviation about the mean in either direction (minimum 10 data points) at a given temperature. The mismatch between $\alpha_\mathrm{TGS}$ of Au-coated SiC and $\alpha_\mathrm{LFA}$ can be attributed to poor fits caused by the reduced TGS signal intensity after the first SAW, as shown in Figure \ref{fig:Au_vs_W_coat}. Because the numerical fitting used only the portion of the signal (fitted signal) after the first SAW was completed, the reduced intensity of the fitted signal degraded the fit quality for the Au-coated samples SiC\cite{dennett_thermal_2018}. This is consistent with larger error bars observed for $\alpha_\mathrm{TGS}$ of Au-coated 3C-SiC.

Despite improved fits and reduced errors, $\alpha_\mathrm{TGS}$ of W-coated SiC is lower than $\alpha_\mathrm{LFA}$. In contrast to the Au-coated SiC case, this difference is not a fitting artifact but arises from the suppression of phonons with MFPs on the order of or larger than $\Lambda$, where $\Lambda$ is the thermal grating period of the TGS measurement. 
This effect is discussed in detail in the subsequent sections \cite{dennett_capturing_2019, huberman_unifying_2017, maznev_onset_2011}. 

Similarly, Figure \ref{fig:temperature_dep_alpha_and_SAW} (b) shows that the Young's modulus (E) of W-coated SiC measured using TGS, when normalized by the room temperature E, agrees with theory (refer to the Supplementary Material for the calculation of the theoretical value). However, the inset in Figure \ref{fig:temperature_dep_alpha_and_SAW} (b) shows that there is a difference between the absolute values. Given that the SAW-speed difference is $< 5 \%$ and the theoretical E values exhibit substantial scatter (see Supplementary Material), we consider this difference to be small. Also, note that $E$ of Au-coated SiC has not been reported since this could not be reliably extracted by computing the power spectral density (PSD) of the TGS signal.

Given the difficulty in fitting the TGS signal acquired using Au-coated 3C-SiC, the associated errors in $\alpha_\mathrm{TGS}$ and the inability to extract E, W-coated 3C-SiC was used for the remainder of this work. 

\subsection{Capturing changes in thermal diffusivity of CVD 3C-SiC during ion irradiation using TGS}\label{ref:thermomech_vs_dose}

Thermal diffusivity and SAW speed of CVD 3C-SiC were measured as a function of dose up to approximately 1 dpa during Si$^{4+}$ ion irradiation at 300\degree C (S1 and S2), 525\degree C (S4) 550\degree C (S3) using TGS, as shown in Figure \ref{fig:thermal_diffusivity_vs_dose_v2} and Figure 3 in the Supplementary Material. The measured thermal diffusivity values are normalized by the pre-irradiation averages (Table \ref{tab:saturation_thermal_diff}), obtained from ten measurements prior to irradiation. This normalization was done because $\alpha_\mathrm{TGS}$ of 3C-SiC was found to be systematically lower than $\alpha_\mathrm{LFA}$ as discussed in Section \ref{sec:thermomechanical_temp}. However, when normalized by a consistent baseline, such as room temperature $\alpha_\mathrm{TGS}$, the results were in agreement with those of $\alpha_\mathrm{LFA}$ as shown in Figure \ref{fig:temperature_dep_alpha_and_SAW}. Accordingly, throughout the remainder of this paper $\alpha_\mathrm{TGS}$ is not reported as an absolute value, but is instead reported as a normalized value and thus used to infer relative changes.

\begin{figure}
    \centering
    \includegraphics[width=0.75\linewidth]{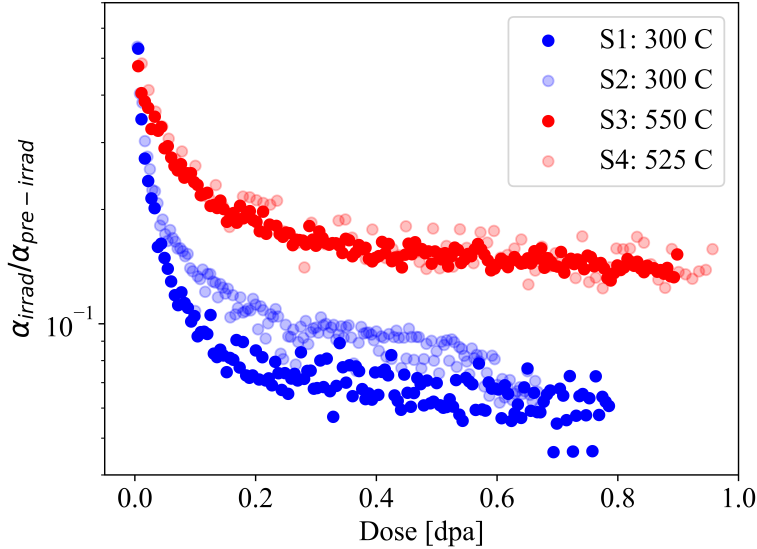}
    \caption{Normalized thermal diffusivity of CVD 3C-SiC as a function of dose at 300, 525 and 550 \degree C as measured using TGS.}
    \label{fig:thermal_diffusivity_vs_dose_v2}
\end{figure}

It is well established that $\alpha$ of 3C-SiC drops sharply after exposure to fission neutrons or ions, with the drop in $\alpha$ being inversely related to irradiation temperature owing to the Arrhenius nature of defect mobility \cite{snead_limits_2004, snead_thermal_2005, snead_handbook_2007, katoh_stability_2011, liu_thermal_2025}. As can be seen in Figure \ref{fig:thermal_diffusivity_vs_dose_v2}, a similar trend was observed for $\alpha_\mathrm{TGS}$ of 3C-SiC during ion irradiation where a sharp drop was seen at low dose, followed by a transition to a more gradual change. The irradiations were repeated at similar temperatures, on different samples, using different thermocouples. Despite the different setups, similar trends in the dose dependence of thermal diffusivity was seen during ion irradiation under similar conditions. This shows the repeatability of ion irradiation and TGS in terms of capturing irradiation induced changes to thermal diffusivity of CVD 3C-SiC despite the sources of error inherent to ion irradiation (see Supplementary Material for a discussion on these). Note that in Figure \ref{fig:thermal_diffusivity_vs_dose_v2} the sudden decrease in $\alpha$ for S2 after 0.6 dpa is an experimental artifact due to sample movement and consequent optics re-alignment. As such, all data for S2 beyond 0.6 dpa are excluded from the analysis for the remainder of this paper. 

Furthermore, it is also well established that the dose at which the drop in $\alpha$ of 3C-SiC shifts to an apparent saturation, also referred to as the saturation dose, increases with irradiation temperature, consistent with point-defect kinetics \cite{katoh_stability_2011, was_fundamentals_2017}. By fitting
the data presented in Figure \ref{fig:thermal_diffusivity_vs_dose_v2} to a piecewise function, where the first region captures the rapid drop while the second captures the `leveling off', the saturation doses for S1, S2, S3 and S4 were found to be 0.15, 0.17, 0.25, 0.30 dpa respectively. Thus, consistent with neutron irradiation data, the ion-irradiation data also showed an increase in saturation dose with temperature. 
Interestingly, in our measurements, we did not observe a saturation at either temperature. This can possibly be explained by the rate theory model of point defect kinetics according to which the defect saturation stage (vacancy concentration does not change with time) follows the stage where the time-dependence of defect concentration is driven by the loss of interstitials to sinks (vacancies concentration increases with time) \cite{was_fundamentals_2017}. As interstitials are lost to sinks, a smaller fraction of interstitials are available to recombine with vacancies, leading to an increase in vacancy concentration. This increase is likely represented in the sharp drop in $\alpha$ seen in Figure \ref{fig:thermal_diffusivity_vs_dose_v2}. It is unlikely that the first and second stages of point defect kinetics, namely the defect build-up and recombination dominated stages, are captured since these on the order of seconds while the time resolution of our system is 60 s \cite{martinez_point_2020}). Then, the behavior seen here is likely the transition from the interstitial sinking stage to the saturation stage. Also, note that the slopes are enhanced, particularly for S1 and S2, given the log scale on the y axis. This represents a practical benefit of using TGS as an \textit{\textit{in situ}} tool since the fluence dependence of properties is measured on a single sample with control over irradiation temperature and dose rate, limiting sample-to-sample variability. Since the decrease in thermal diffusivity after $\sim$ 0.1 dpa is small for all irradiation temperatures, an ex-situ study using separate samples might infer saturation \cite{katoh_stability_2011, liu_thermal_2025}. Refer to the Supplementary Material for changes in a SAW speed as a function of dose.  

\subsection{Pre- and post-irradiation microstructure}\label{sec:microstructure}

\begin{figure}[h!]
    \centering
    \includegraphics[width=\linewidth]{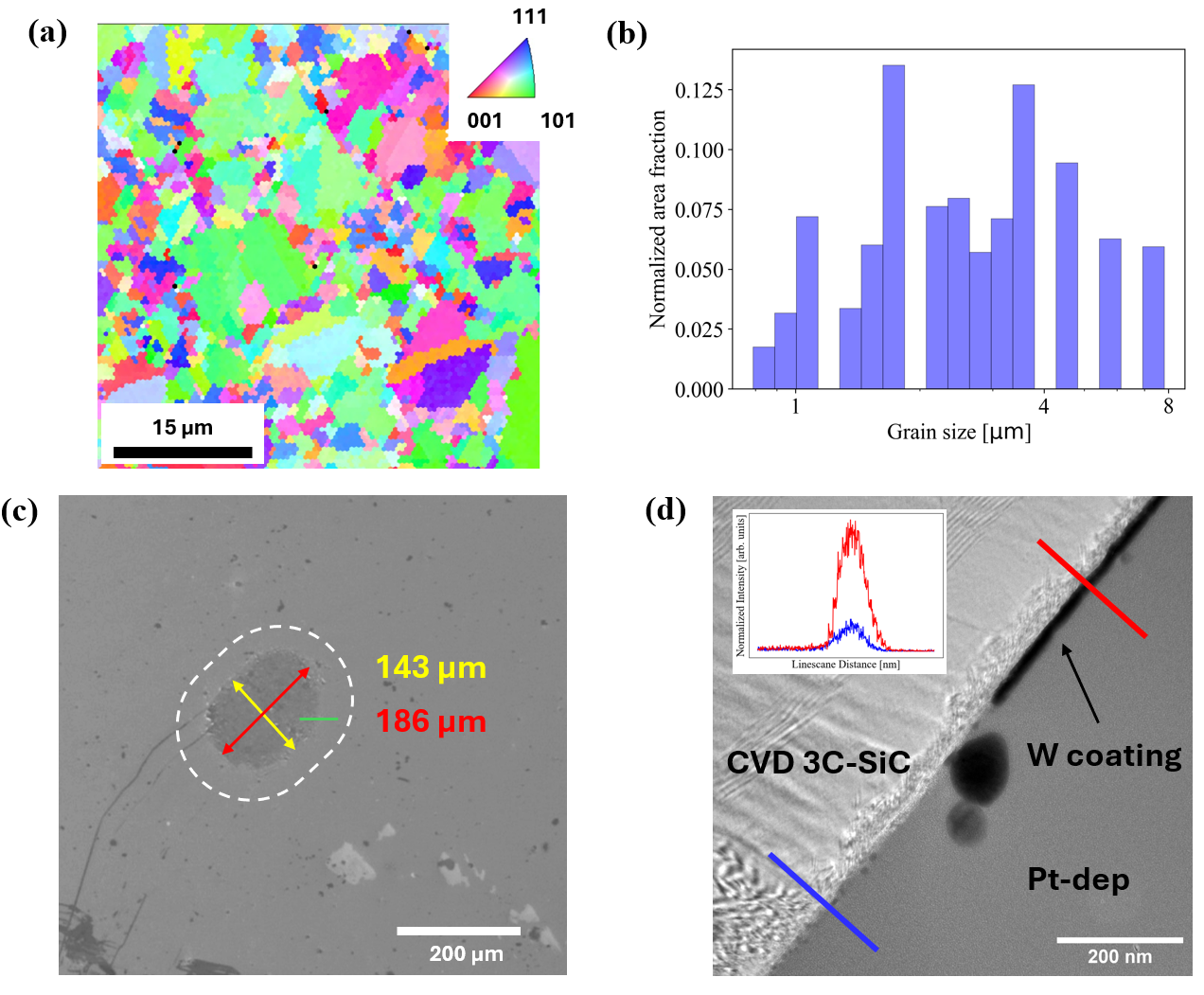}
    \caption{(a) EBSD map showing surface orientation of grains (b) Bar plot showing normalized areal coverage as a function of grain size. (c) SEM micrograph showing the tungsten delamination region (darker region inside white-dashed ellipse), which aligns with the TGS spot on S3. Green marking shows approximate region where TEM lamella was milled (d) TEM micrograph showing cross section of boundary between the region where the tungsten coating had delaminated and where it was intact. Blue and red lines mark approximate locations of EDS line-scans, with the inset showing tungsten signal intensity.}
    \label{fig:microstructure}
\end{figure}

The EBSD map and grain size distribution for unirradiated CVD 3C-SiC are presented in Figure \ref{fig:microstructure}. The Kikuchi patterns at all points studied correspond to those of 3C-SiC, with a relatively random distribution in grain orientations and sizes. Although the manufacturer specified an average grain size of $5 \; \mathrm{\upmu m}$, the area-weighted average grain size seen in the EBSD map was $3 \pm 1.8$ $\mathrm{\upmu m}$. The grain size distribution shows some grains nearly twice as large as the average. 

Four CVD-SiC samples (S1-S4) were coated with W prior to Si$^{4+}$ irradiation to 0.5-1 dpa (see Section \ref{sec:Methods}, Table \ref{tab:saturation_thermal_diff}). The tungsten coating on S3 (sample exposed to highest dose and temperature) was examined by TEM and SEM following irradiation. The TEM lamella was prepared from within the ion beam spot and spanned a region where the tungsten coating had delaminated and a region where it remained intact. As shown in Figure \ref{fig:microstructure}(c), the delamination area is roughly circular and measures $\sim  150 \: \mathrm{\upmu m}$ across, comparable in size to the pump laser spot \cite{wylie_accelerating_2025}. This suggests that the combined energy of the TGS laser and the ion beam (incident at 45\degree) might have contributed to sputtering or delamination of the tungsten coating. Similar delamination features were also observed using optical microscopy in samples irradiated at 300\degree C, whereas samples not subjected to simultaneous laser and ion beam exposure showed no delamination. The TEM micrograph and the associated EDS line scans in Figure \ref{fig:microstructure} (d) confirm the presence of tungsten in the region where the tungsten coating was intact, whereas the reduced intensity of the tungsten signal confirms the likely removal of the tungsten coating in the delamination region. 

In the region where the tungsten coating remained intact, its thickness was approximated by evaluating multiple intensity profiles across the tungsten layer. The average thickness was $19.2 \: \pm \: 1 \: \mathrm{nm}$. The process of extracting the thickness is described in the Supplementary Material.

\subsection{Length-scale dependence of thermal conductivity of single crystal and polycrystalline 3C-SiC}\label{sec:length_scale_dep}

First principles DFT+BTE simulations were performed for single crystal and polycrystalline 3C-SiC, with the polycrystal defined by a nominal grain size of $L_{gb} \: = \: 3.5 \: \mathrm{\upmu m}$. DFT+BTE simulations discussed below were performed at 27 and 527\degree C. The effective thermal conductivity ($K_{eff}$) is defined as the thermal conductivity predicted at a given thermal grating period ($\Lambda$) normalized by the corresponding bulk value. The cumulative thermal conductivity is defined as the thermal conductivity accumulated as a function of phonon MFP, normalized by the thermal conductivity of the complete phonon spectrum. Thermal conductivity and thermal diffusivity are related as $K = \rho C_p\alpha$, where $K$ is the thermal conductivity, $\rho$ is the mass density, $C_p$ is the specific heat capacity and $\alpha$ is the thermal diffusivity. Because $C_p$ and $\rho$ are insensitive to the length scales considered here, changes in K are considered to map directly onto changes in $\alpha$ \cite{cheng_high_2022}. 

\begin{figure}[h!]
    \centering
    \includegraphics[width=\linewidth]{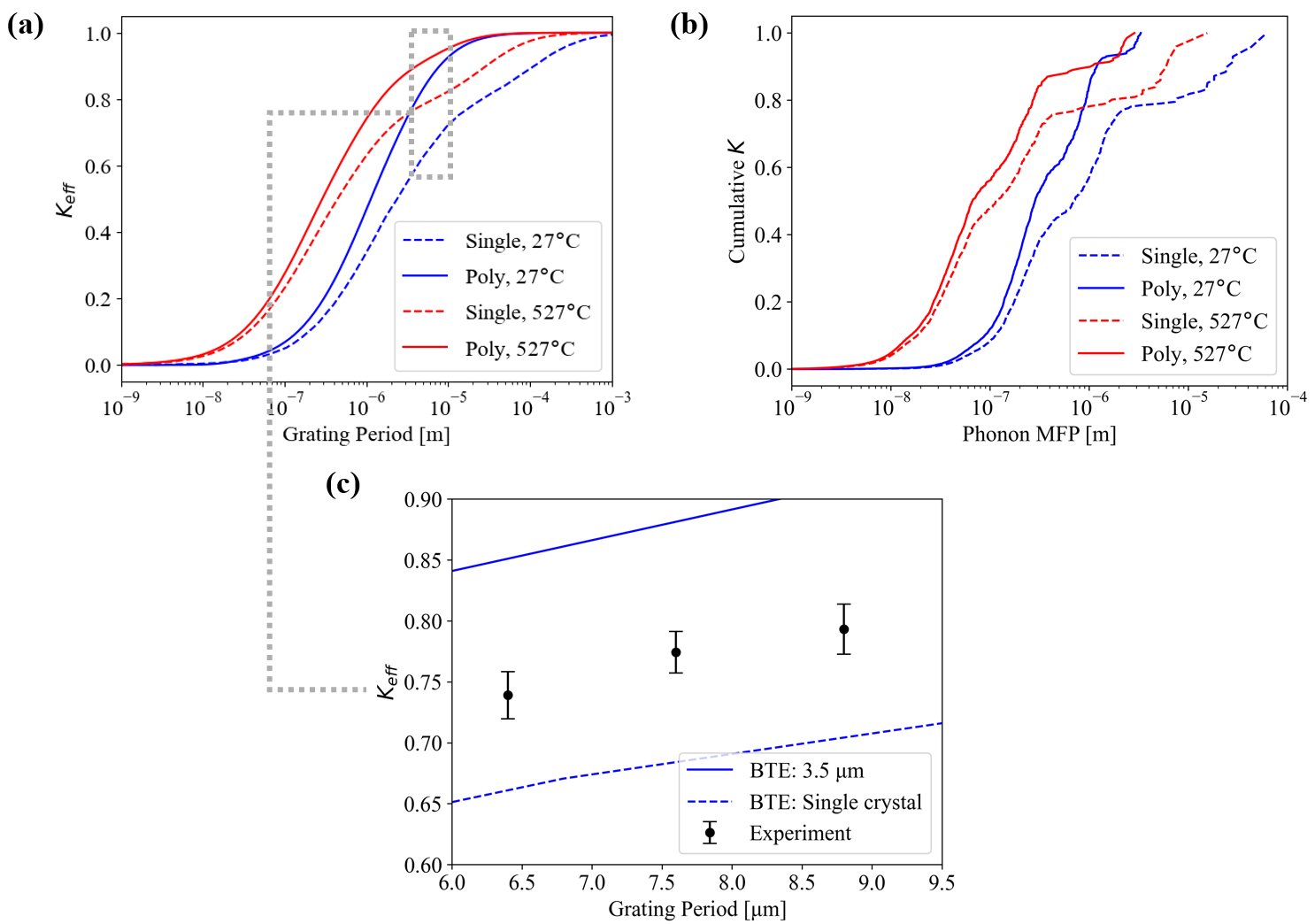}
    \caption{(a) DFT+BTE predictions for $K_{eff}(\Lambda)$ for pristine single crystal and polycrystalline 3C-SiC calculated at 27 and 527\degree C. (b) Spectral contribution to cumulative thermal conductivity (normalized) for single and polycrystalline 3C-SiC. (c) Zoomed in view for $K_{eff}(\Lambda)$ bounded between $\Lambda \: = \: 6 \: \mathrm{\upmu m}$ and $\Lambda \: = \: 9.5 \: \mathrm{\upmu m}$ at 27\degree C. Also shown are experimental results for $K_{eff}(\Lambda)$ of CVD 3C-SiC taken at nominal $\Lambda \: = \: 6.4$, $7.6$ and $8.8 \: \mathrm{\upmu m}$. The experimental data was acquired at room temperature (between 20 and 22\degree C).}
    \label{fig:keff_vs_grating_pure_nograins_wgrains}
\end{figure}

Figure \ref{fig:keff_vs_grating_pure_nograins_wgrains}(a) shows that $K_{eff}$ increase with $\Lambda$ and approaches unity, consistent with prior reports for Si, SiGe alloys, and 3C-SiC \cite{dennett_thermal_2018, huberman_unifying_2017, cheng_high_2022, anufriev_nanoscale_2022}. Introducing grain boundaries shifts the $K_{eff}(\Lambda)$ curve upward, bringing it closer to unity at smaller $\Lambda$. The cumulative thermal conductivity vs phonon MFP curve provides the spectral contribution to thermal conductivity. As shown in Figure \ref{fig:keff_vs_grating_pure_nograins_wgrains} (b), in defect-free single crystal 3C-SiC, phonons with mean free paths (MFPs) spanning $\sim 10^{-8}$ m to  $6 \times 10^{-5}$ m contribute to the thermal conductivity, whereas in defect-free polycrystalline 3C-SiC the contributing long-MFP range is truncated at $\sim 3 \times 10^{-6}$ m. As the temperature increases to 527\degree C the $K_{eff}(\Lambda)$ curves are displaced leftward. The spectral contribution, is similarly shifted left for both single and polycrystalline 3C-SiC. This can be explained by the difference in grain boundary and Umklapp scattering. Grain boundary  scattering ($\tau_{gb} \: = \: v_g/L_{gb}$ where $\tau_{gb}$ is the time a phonon with frequency-dependent group velocity, $v_g$, travels between two consecutive grain boundary scattering events) was treated such that, in accordance with Matthiessen's rule (Eq. \ref{eq:scattering_rates}), high frequency phonons with intrinsically shorter MFPs are less affected while lower frequency phonons with intrinsically longer MFPs are more strongly affected. As a result, long MFP phonons are truncated to the grain size. However, Umklapp scattering was frequency dependent ($\tau_U \: \propto \omega^{-2}$ where $\tau_U$ is the time a phonon with frequency $\omega$ travels between two consecutive anharmonic scattering events) and so all phonons experienced reductions in MFP without a sharp truncation bound \cite{snead_thermal_2005}.

In Figure \ref{fig:keff_vs_grating_pure_nograins_wgrains} (b), it is evident that phonons with MFPs spanning many orders of magnitude contribute to the thermal conductivity of 3C-SiC at 27 and 527\degree C. The experimental length scales for TGS are determined by $\Lambda$. Figure \ref{fig:keff_vs_grating_pure_nograins_wgrains} (a) shows that $K_{eff}$ increases with $\Lambda$. This can be attributed to the suppression of phonons with MFP on the order of and larger than $\Lambda$. So, when $\Lambda$ increases, fewer phonons are suppressed, leading to a higher $K_{eff}$. The mathematical formulation for this suppression was provided elsewhere, but is summarized hereafter \cite{maznev_onset_2011}. In techniques such as TGS, two "pump" laser beams interfere on a surface to create a periodic "hot and cold" temperature grating (hot and cold regions are spaced by $\Lambda/2$). The grating relaxes as the laser energy that is deposited in the hot regions is transferred to the cold regions. In ceramic materials like 3C-SiC, this transfer of thermal energy is dependent on phonons. When the thermal grating period (1-10 $\mathrm{\upmu m}$ typically used in TGS when combined with ion irradiation) approaches the MFP of phonons, the longest MFP phonons traverse multiple hot and cold regions before scattering or equilibrating. As a result, long MFP phonons transport energy over long distances but do not exchange it efficiently on the short length scale needed to smooth the periodic modulation. As such, their contribution to grating relaxation is suppressed relative to their contribution to bulk heat conduction. The decay is therefore slower than for the bulk. Increasing $\Lambda$ progressively restores the bulk limit, as shown in Figure \ref{fig:keff_vs_grating_pure_nograins_wgrains}, as more of the phonon spectrum re-equilibrates within $\Lambda/2$. The phonon population can then be bounded by two extremes that are identified by $ql$, where $q = 2\pi/\Lambda$ is the grating wave-vector and $l$ is the phonon MFP. When $ql \ll 1$, phonon MFPs are much smaller than the thermal grating period and their contribution to relaxing the temperature grating will not be suppressed. These are referred to as diffusive phonons. However, when $ql \gg 1$, phonons can be considered as ballistic, or non-diffusive, and their contribution will be significantly suppressed. The suppression as a function of phonon MFP is plotted in the Supplementary Material. Grain boundaries act as periodic scattering sites, truncating the long-MFP tail present in single crystal 3C-SiC (effectively limiting $l_\mathrm{max} \: \leq \: L_\mathrm{gb}$). This reduces the fraction of heat carried by the ballistic phonons that are suppressed in TGS, yielding $K_{eff}$ values closer to the bulk limit at a given $\Lambda$. 

To experimentally validate the predicted $\Lambda$ dependence, $\alpha_\mathrm{TGS}$ of CVD 3C-SiC was measured at $\Lambda \: = \: 6.4$, $7.6$ and $8.8 \: \mathrm{\upmu m}$ under vacuum ($< \: 5 \times10^{-5}$ Torr) at approximately 20 - 22\degree C. For each $\Lambda$, $\alpha_\mathrm{TGS}$ was measured ten times at ten distinct locations on the same sample (100 total measurements at each $\Lambda$). Repeated measurements at each location were first averaged, and the reported mean was computed from these location-averaged values.  Then, the effective thermal conductivity, $K_{eff}(\Lambda)$, was defined as $K_{eff}(\Lambda) \: = \: \alpha_\mathrm{TGS}(\Lambda)/\alpha_\mathrm{LFA}$, where $\alpha_\mathrm{LFA}$ was measured at 25\degree C. The resulting $K_{eff}(\Lambda)$ values are shown in Figure \ref{fig:keff_vs_grating_pure_nograins_wgrains}(b) along with predictions from DFT+BTE simulations for single and polycrystalline 3C-SiC at 27\degree C; raw data for $K_{eff}$ at each $\Lambda$ are provided in the Supplementary Material. Because the experimentally accessible range of $\Lambda$ is narrow compared to the breadth of the phonon MFP spectrum, the change in $K_{eff}$ is modest. Therefore statistical analysis (presented in Supplementary Material) of the experimental results was performed, confirming that $K_{eff}$ increases with $\Lambda$ and that the trend is distinguishable from measurement scatter.

\subsection{Length-scale dependence of thermal conductivity of vacancy-rich 3C-SiC}\label{sec:length_scale_vacancy}

DFT+BTE simulations were also performed for single crystal and polycrystalline 3C-SiC as a function of vacancy concentration at various thermal grating periods (or spacings), $\Lambda$. The bulk thermal conductivity values at 27 and 527\degree C as a function of vacancy concentration are listed in the Supplementary Material. 
As described in Section \ref{sec:Methods}, vacancy scattering was calculated based on $\epsilon$ values of 1 and 10. $\epsilon$ is a fitting parameter that is used to relate the size mismatch between a point defect and the host species on the same sublattice to the corresponding change in force constants, with typical values between 1 and 300. 
$\epsilon \: = \: 10$ was found to provide reasonably good agreement with prior atomistic predictions based on the heat-current correlation (Green-Kubo) approach in molecular dynamics \cite{gurunathan_analytical_2020, li_atomistic_1998, samolyuk_molecular_2011}. However, while computational results presented by Li et al. showed that the thermal conductivity of defect-rich 3C-SiC is temperature insensitive, our results show temperature sensitivity. This could be explained by the difference in how lattice thermal conductivity is treated by BTE and MD \cite{wei_influence_2023} and is discussed in detail in the Supplementary Material.

\begin{figure}[h!]
    \centering
    \includegraphics[width=\linewidth]{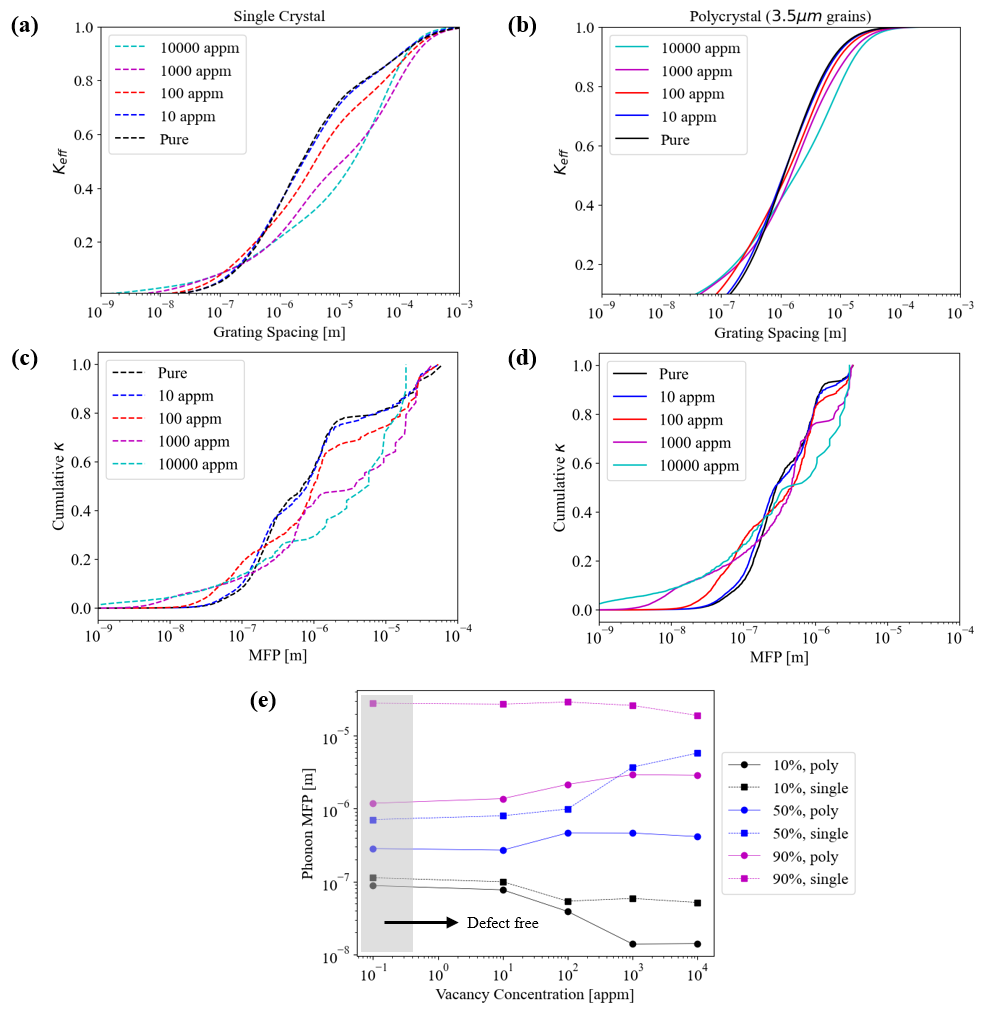}
    \caption{DFT+BTE predictions for $K_{eff}(\Lambda)$ as a function of vacancy concentration for single crystal (a) and polycrystalline 3C-SiC with $3.5 \: \upmu m$ grains (b) at 27\degree C. Also shown, for reference, are the $K_{eff}(\Lambda)$ curves for the pure material. Cumulative thermal conductivity vs phonon MFP for single (c) and polycrystalline (d) 3C-SiC. $K_{irr}/K_0$ shown for single crystal. (e) Phonon MFP at which cumulative thermal conductivity reaches 10, 50 and 90\% plotted versus vacancy concentration for single and polycrystalline 3C-SiC.}
    \label{fig:keff_vs_grating_spacing_final}
\end{figure}

$K_{eff}(\Lambda)$ for single crystal and polycrystalline (3.5 $\upmu$m grains) 3C-SiC at 27\degree C with 0 (defect-free or pristine), 10, 100, 1000 and 10000 appm vacancies is shown in Figure \ref{fig:keff_vs_grating_spacing_final} (a) and (b). Compared to the pristine case, the effect of vacancies on the length scale required to reach $K_{eff} \: \sim \: 1$ is small. In contrast, length scales required to reach $K_{eff} \: \geq \: 0.7$, increase with vacancy concentration, with the difference being larger for single crystal 3C-SiC as compared to polycrystalline 3C-SiC. This means that at a given $\Lambda$, especially between 1 and 10 $\mathrm{\upmu m}$ the difference between $\alpha$ measured using TGS and bulk techniques grows with increasing vacancy concentration. 

The cumulative thermal conductivity versus phonon mean free path (MFP) (Figure \ref{fig:keff_vs_grating_spacing_final} (c) and (d)) provides insight into the behavior described above. For all vacancy concentrations, the MFP at which 10\% of the conductivity is accumulated shifts to smaller values, with a larger shift as the vacancy concentration increases (Figure \ref{fig:keff_vs_grating_spacing_final}(e)). This is consistent with Rayleigh (point-defect) scattering, $\tau_{\mathrm{vac}} \propto \omega^{-4}$, according to which high-frequency phonons that already have relatively short intrinsic MFPs are preferentially scattered\cite{klemens_theory_1984}. At the same time, the MFP at which 50\% of the conductivity is accumulated shifts to larger values followed by a sharper rise in the accumulation curve. This trend is more evident for single crystal 3C-SiC as compared to polycrystalline 3C-SiC. For single-crystal 3C-SiC with $\geq 1000$ appm vacancies at 27\degree C, a substantial fraction of the remaining conductivity accumulates over a relatively narrow MFP band ($10^{-6}$ m - $10^{-5}$ m). Thus, adding vacancies broadens the contribution from short-MFP phonons (more of the intrinsically shorter MFP spectrum contributes), while effectively narrowing the long-MFP contribution (fewer low-frequency modes dominate what remains). In the polycrystalline material, this trend is weaker because grain boundaries truncate the long-MFP tail. Increasing the temperature from 27\degree C to 527\degree C shifts the accumulation curves to smaller MFPs, but the qualitative length-scale dependence remains: larger $\Lambda$ yield higher $K_{\mathrm{eff}}$, and the strength of this dependence increases with vacancy concentration, especially for $\Lambda  \: = \: 1  \: - \: 10 \: \mathrm{\upmu m}$. Plots showing this trend for 527\degree C are provided in the Supplementary Material.

\section{Discussion}

\subsection{Effect of reflective coating on TGS signal}\label{sec:reflective_coating}

It was shown in Section \ref{sec:reflective_coating_results} that the use of Au and W reflective coatings on 3C-SiC permits the optical study of surface gratings using TGS. However, the effect of the coating on the measured properties should be understood. As discussed by K\"ading et al., the effect of the coating can be considered in three limiting cases based on a $qL_{coat}$ factor, where $q = 2\pi/\Lambda$ is the grating wavevector and $L_{coat}$ is the thickness of the coating \cite{kading_transient_1995}. For the $qL_{coat} \ll 1$ case, it was shown that the effect of the coating on the measured properties of the bulk can be ignored. Given a coating thickness of $20 \; \mathrm{nm}$ (Figure \ref{fig:microstructure}) and a $\Lambda = 6.4 \: \mathrm{\upmu m}$, $qL_{coat} = 0.016$. Therefore, for all results discussed in this work, it can be assumed that the reflective coating does not have an impact on the properties measured using TGS. A possible exception to this assumption is when the thermal diffusivity of the coating is substantially ($\sim 100 \times$) higher than that of the substrate. In the present case, it can be assumed that $\alpha$ of the W layer is less than $50 \: \mathrm{mm^2/s}$ while that of the irradiated CVD 3C-SiC is approximately $2 \: \mathrm{mm^2/s}$ as shown in Table \ref{tab:saturation_thermal_diff} \cite{wylie_accelerating_2025}. This difference is smaller than the two order magnitude difference reported by K\"ading et al. \cite{kading_transient_1995}. Furthermore, as shown in Section \ref{sec:microstructure}, the W coating was likely lost during the combined exposure to the laser and ion beams. Therefore, it is possible that this point need not be considered. 

Thus, although the absorption coefficient for visible light of high-purity CVD 3C-SiC is low, making it hard to obtain a TGS signal without a reflective coating, the introduction of defects through ion irradiation likely increases it, allowing acquisition of TGS data despite the removal of the W reflective coating due to combined ion irradiation and laser exposure. This is consistent with similar results reported in literature which have shown that the addition of defects, such as vacancies, increases the absorption of visible light by SiC and Al$_2$O$_3$ \cite{zhang_influence_2024, wendler_damage_2012, lin_origins_2025}. This has been attributed to production of valence bands in the electronic band structure that reduce the bandgap, allowing materials like 3C-SiC to interact with the pump (532 nm) and probe (577 nm) lasers used in this study. Kukushikin et al. used first principles calculations to estimate that, for 1 and 3.7 at.\% negatively charged Si vacancies, the optical penetration depth of 532 nm laser in 3C-SiC is approximately 65 nm and 63 nm, respectively \cite{kukushkin_dielectric_2022}. They also experimentally demonstrate that the penetration depth of 532 nm light in as-grown 3C-SiC is nearly infinite, while the addition of thermal vacancies reduces this to 162 nm. Similarly, Zhang et al. used DFT to estimate penetration depths of approximately 38 nm and 92 nm for 6.25 at.\% Si and C vacancies in 3C-SiC, respectively \cite{zhang_influence_2024}. These results confirm that vacancies strongly reduce the penetration depth of 532 nm light in 3C-SiC, supporting our experimental observation that TGS can be performed on defect-rich 3C-SiC without applying reflective coatings.

Although we argue that the effect of the reflective coating can be ignored in the present study, future work should consider solving the multi-layer transient thermal grating equations to extract the thermal diffusivity of each layer and the interfacial resistance from a single measurement \cite{kading_transient_1995}. This would separate the effects of the reflective coating from those of the substrate while also helping accelerate the development and evaluation of functional coatings.

\subsection{Effect of grain size and inhomogeneity on thermal diffusivity measurements of 3C-SiC using TGS}\label{sec:grain_inhomogeneity}

Figure \ref{fig:keff_vs_grating_pure_nograins_wgrains} shows that the experimentally measured $\alpha_\mathrm{TGS}$ falls between the computational predictions for single crystal 3C-SiC and polycrystalline 3C-SiC having $L_{gb} \: = \: 3.5 \: \mathrm{\upmu m}$. This result can likely be explained by the grain size distribution of CVD 3C-SiC. As shown in Figure \ref{fig:microstructure} (a) and (b), CVD 3C-SiC has grains larger than $3.5 \: \mathrm{\upmu m}$ leading to regions of reduced phonon-grain boundary scattering and decreased $K_{eff}$. However, for the purpose of this study, this simplification was deemed sufficient to show that the addition of grain boundaries increases $K_{eff}$ at a given $\Lambda$ when compared to single crystal 3C-SiC.

These results indicate that $\alpha_\mathrm{TGS}$ can converge to $\alpha_\mathrm{LFA}$ in phonon-dominated materials when the experimental length scales are large compared to the MFPs of the dominant heat-carrying phonons (i.e. $ql \ll 1$ across the relevant spectrum). Liu et al. demonstrated the agreement between $\alpha_\mathrm{TGS}$ and $\alpha_\mathrm{LFA}$ for sintered SiC with sub-100-nm grains using $\Lambda \: = 5.1 \: \mathrm{\upmu m}$, where grain-boundary scattering truncates long-MFP contributions and the TGS measurement recovers bulk-like diffusivity \cite{liu_thermal_2025}. However, although they conclude that TGS provides an accurate measure of $\alpha$ of SiC-based materials, our results show that is microstructure dependent and not universally applicable. More generally, when the experimental length scale is comparable to, or smaller than, the upper end of the heat-carrying MFP spectrum (such as in materials with grain sizes on the order of or larger than $\Lambda$), TGS should be interpreted as measuring an effective, length-scale-dependent diffusivity rather than a bulk value. In such cases, reporting diffusivity relative to a reference state (e.g., room-temperature or pre-irradiation $\alpha$) can be used as a metric for tracking irradiation or processing-induced changes.


\subsection{Difference between point defect induced thermal diffusivity degradation measured using TGS and bulk techniques}\label{subsec:grating_spacing_thermal_diff}

\begin{figure}[h!]
    \centering
    \includegraphics[width=\linewidth]{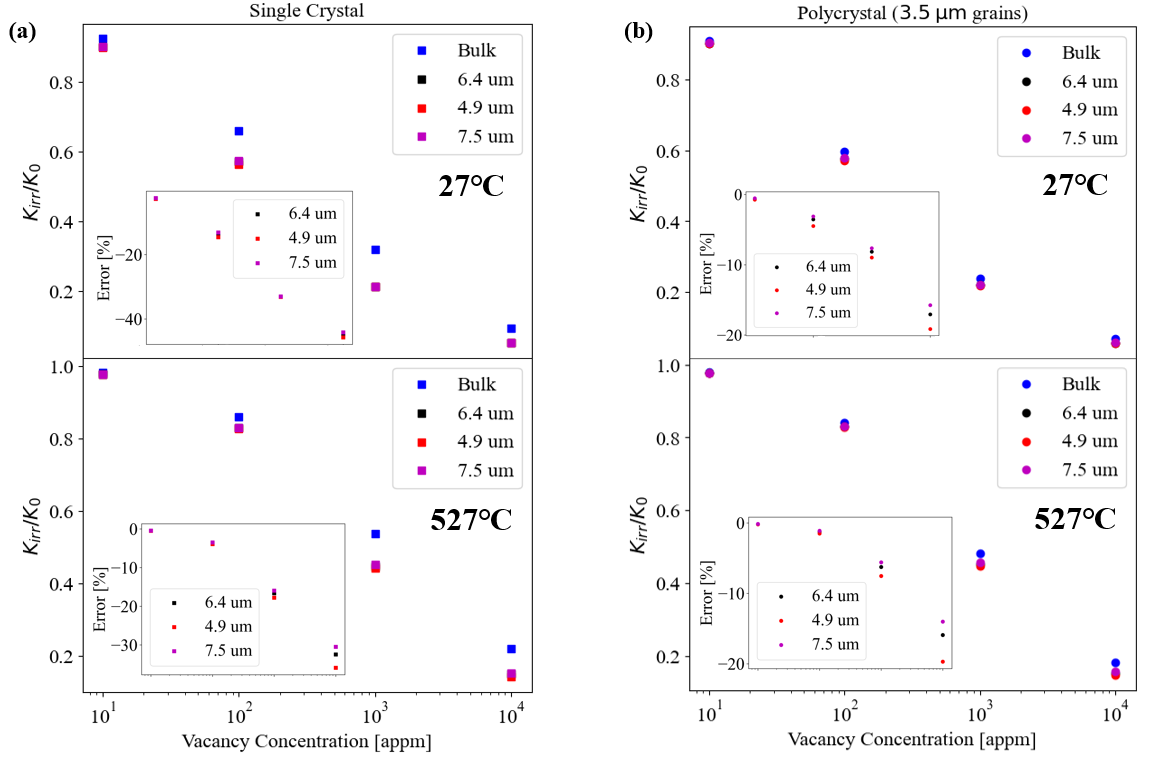}
    \caption{DFT+BTE predictions for thermal conductivity degradation ($K_\mathrm{irr}/K_0$) as a function of vacancy concentration for (a) single and (b) polycrystalline 3C-SiC. Results are presented for 27 and 527\degree C. Insets in the plot show the error between bulk and TGS measurements following the data presented in the main figure.}
    \label{fig:kirr_k0}
\end{figure}

A central motivation of this work is to determine whether TGS can quantitatively track irradiation-induced degradation in thermal conductivity (or diffusivity) in ceramics. We assess this by first comparing the computationally predicted conductivity degradation, $K_{\mathrm{irr}}/K_0$, as a function of vacancy concentration for single-crystal and polycrystalline 3C-SiC (Figure \ref{fig:kirr_k0}). Here $K_0$ is defined as the defect-free conductivity evaluated at the same experimental length scale (i.e., the same $\Lambda$) as $K_{\mathrm{irr}}$.

Across both microstructures and at 27 and 527\degree C thermal conductivity drops with increasing vacancy concentration in both TGS-relevant and bulk length scales (Figure \ref{fig:kirr_k0}). In this section, we focus on $\Lambda \: =$ 4.6, 6.4 an 7.5 $\mathrm{\upmu m}$, which span the typical range available during simultaneous ion irradiation and TGS measurements (as a consequence of the shallow implantation depth of ions, $\Lambda \leq 10 \: \mathrm{\upmu m}$ must be used). The analysis presented here can be extended to any $\Lambda$ between $10^{-9}$ and $10^{-3}$ m using the raw data provided. Note that in Figure \ref{fig:kirr_k0}, `Bulk' refers to the smallest $\Lambda$ at which $K_{eff} \: = \: 1$, representing the length scales that are expected to yield an experimental results quantitatively equivalent to LFA. While both TGS and bulk measurements yield a monotonic degradation in thermal diffusivity with increasing vacancy concentration, the magnitude of the predicted degradation differs between the two for single and polycrystalline 3C-SiC and at both temperatures. This is important because in Section \ref{sec:thermomechanical_temp} it was shown that the trend in $\alpha$ with increasing temperature acquired by TGS and LFA are quantitatively equal when $\alpha$ is normalized by the corresponding room temperature value. Motivated by this result, irradiation-induced degradation in Figure \ref{fig:thermal_diffusivity_vs_dose_v2} was reported after normalizing by the pre-irradiation average. However, the results in Figure \ref{fig:kirr_k0} suggest that beyond qualitatively capturing trends, the quantitative mapping from TGS to bulk-relevant thermal degradation warrants further investigation.

In Section \ref{sec:length_scale_vacancy} we show that the spectral contribution to thermal conductivity evolves with vacancy concentration. At low vacancy concentrations, this change is small. As the vacancy concentration increases, there is a pronounced shift in the spectral contribution due to the strong scattering of high-frequency (intrinsically short-MFP) phonons, which increases the relative contribution of lower-frequency, longer-MFP modes. As discussed in Section \ref{sec:length_scale_dep} the contribution of phonons to the relaxation of the temperature grating is suppressed when the length scales of the phonons are equivalent to or larger than those of the grating. Therefore, as the vacancy concentration increases, the spectral contribution of the long-MFP phonons to $\alpha$ increases. Consequently, TGS will exaggerate the degradation of $\alpha$ with increasing vacancy concentration, leading to larger errors between the degradation measured by TGS and bulk techniques as shown in the insets in Figure \ref{fig:kirr_k0}. Similar effects are seen in polycrystalline 3C-SiC, although to a lesser extent given the truncation of long MFP phonon tail through the grain boundaries. Thus, while the grain boundaries work to reduce the effect of length scales, vacancies work to enhance it by biasing the spectral contribution towards long MFP modes and therefore longer length scales. As a result, $K_{eff}(\Lambda)$ deviates from the pure case with increasing vacancy concentration, but to a lesser extent than in single crystal 3C-SiC (Figure \ref{fig:keff_vs_grating_spacing_final} (a) and (b)). Furthermore, for both single and polycrystalline 3C-SiC, the magnitude of the error between degradation measured by TGS and bulk techniques decreases with temperature. This follows from increased Umklapp scattering with temperature which, as explained in Section \ref{sec:length_scale_dep}, induces a leftward shift in the spectral contribution curve and limits the length-scale dependence.

\subsection{Introducing a dimensionless suppression factor}\label{sec:suppression_factor}

Given the relation between thermal conductivity ($K$) and thermal diffusivity ($\alpha$) and knowing that heat capacity and mass density are minimally changed following irradiation induced displacement damage, the following becomes evident \cite{lee_thermal_1982, snead_handbook_2007}: 

\begin{equation}\label{eq:K_irr_and_alpha_irr}
    \frac{K_{\mathrm{irr}}(T)}{K_0(T)} = \frac{\alpha_{\mathrm{irr}}(T)}{\alpha_0(T)}
\end{equation}

\noindent where $T$ is the measurement temperature and must stay the same for the irradiated and un-irradiated measurement. In Section \ref{subsec:grating_spacing_thermal_diff} we argued that as the vacancy concentration increases, $K_{\mathrm{irr}}/K_0$ derived from TGS deviates from that derived using bulk techniques such as LFA. However, we also observed that the extent of this deviation is within 30-40\% for single crystal 3C-SiC and 10-20\% for polycrystalline 3C-SiC (Figure \ref{fig:kirr_k0}). Given that the degradation in $\alpha$ after irradiation is expected to be between one and two orders of magnitude, the following can be argued, especially in the case of polycrystalline CVD 3C-SiC \cite{snead_handbook_2007, katoh_stability_2011, liu_thermal_2025}: 

\begin{equation}\label{eq:alpha_irr_over_alpha_0}
    \frac{\alpha_\mathrm{irr,TGS}(\Lambda,T)}{\alpha_\mathrm{0,TGS}(\Lambda,T)} \sim \frac{\alpha_\mathrm{irr,LFA}(T)}{\alpha_\mathrm{0,LFA}(T)}
\end{equation}

\noindent where $\Lambda$ is the thermal grating period used in the measurement and must remain the same for the pre-irradiated and the irradiated measurements Re-arranging Equation \ref{eq:alpha_irr_over_alpha_0}, the following equation is derived where $S(\Lambda)$ is a dimensionless `Suppression Factor': 

\begin{equation}
    S(\Lambda) = \frac{\alpha_\mathrm{irr,TGS}(\Lambda,T)}{\alpha_\mathrm{irr,LFA}(T)} = \frac{\alpha_\mathrm{0,TGS}(\Lambda, T)}{\alpha_\mathrm{0,LFA}(T)}
\end{equation}

The suppressed contribution of long-MFP phonons due to experimental length scales in techniques like TGS and in thin-film geometries has motivated fully spectral, MFP-dependent corrections when the phonon spectrum is known \cite{maznev_onset_2011, cuffe_reconstructing_2015}. While such treatments are more rigorous, they are not always practical to repeat for every specimen, and irradiation condition. The dimensionless `Suppression Factor', $S(\Lambda)$, could provide a pragmatic alternative for materials with phonon-dominated thermal processes. Once calibrated for a given measurement geometry and material system (crystal structure, expected defect type), it enables TGS measurements of the modified system to be mapped onto bulk-relevant diffusivities using only the pre-modification bulk-reference state. This workflow is broadly applicable to accelerated studies including ion-irradiation campaigns for rapid down-selection of radiation-tolerant ceramics, microstructure engineering strategies (e.g., nanograined materials that increase sink density and reduce steady-state defect concentrations), and dopant/impurity studies in semiconducting ceramics \cite{moriani_point-defect_2006, rose_instability_1997, ivanova_thermoelectric_2006}. In these settings, $S(\Lambda)$ allows quantitative, rapid and \textit{in situ} estimate of bulk transport degradation without requiring full spectral modeling for each condition.

The results above support a practical workflow in which TGS can be used to infer bulk-relevant thermal transport degradation in 3C-SiC through $S(\Lambda)$. However, introducing $S(\Lambda)$ underscores a central argument in this paper: because TGS probes heat flow over a finite length scale set by $\Lambda$, the measured diffusivity should not be interpreted as intrinsically “bulk” without accounting for the length-scale dependence of the underlying thermal transport processes in the material. In the next section, we demonstrate the utility of this approach by using $S(\Lambda)$ to obtain quantitative agreement between the degradation of $\alpha$ of neutron-irradiated CVD 3C-SiC measured using LFA and that of ion-irradiated CVD 3C-SiC measured using TGS.

\subsection{Synergies between thermal diffusivity measurements of neutron and ion irradiated CVD 3C-SiC}

The discussion regarding $S(\Lambda)$ thus far largely has been computational. In addition, while ion irradiation has been used as a proxy for neutron irradiation in some materials under specific circumstances, its general applicability has not been demonstrated. This raises two important questions that need to be answered. Can experiments confirm the DFT+BTE prediction that TGS measurements provide insights into the degradation of $\alpha$, in quantitative agreement with bulk measurements? 
Second, is the degradation of $\alpha$ of CVD 3C-SiC following ion irradiation at 300 and 550\degree C comparable to that observed after neutron irradiation under similar conditions? Through this section, we will attempt to answer these questions. 

In 3C-SiC, below 1000 \degree C, point defects and small clusters dominate the microstructure \cite{snead_handbook_2007, katoh_microstructural_2006, hu_positron_2017}. Given the phonon-dominated thermal diffusivity and the microstructural dominance of point defects in the temperature range studied, calculating the thermal defect resistance would provide insights into the microstructure of 3C-SiC following ion irradiation and how it compares to the microstructure following neutron irradiation. The thermal defect resistance is expressed as \cite{snead_thermal_2005}: 
\begin{equation}\label{eq:thermal_defect_resistance}
    \frac{1}{K_{defect}} = \frac{1}{K_{irr}} - \frac{1}{K_{0}}
\end{equation}

\noindent where $K_{irr}$ and $K_\mathrm{0}$ are the thermal conductivities of the irradiated and unirradiated material, respectively. According to Matthiessen's rule, the total phonon scattering rate can be written as the sum of the scattering rates from individual mechanisms such as point and extended defects, grain boundaries, free surfaces and phonon-phonon interactions (Eq. \ref{eq:scattering_rates}) \cite{lindsay_ab_2013, klemens_theory_1984}. Then, using thermal defect resistance it is possible to quantity the additional resistance introduced by a change in the system, such as the formation of radiation-induced defects or an increase in phonon-phonon scattering with temperature. Thus, if the post-irradiation microstructure is assumed to be dominated by point defects, specifically vacancies, thermal defect resistance can be used to isolate the contribution of these defects to the degradation of $\alpha$. A key advantage of using this quantity is that, following from Matthiessen's rule, it is temperature insensitive as long as $K_\mathrm{irr}$ and $K_{0}$ are measured at the same temperature. Many neutron irradiation studies evaluate material properties at room temperature instead of at the irradiation temperature in order to prevent annealing out defects. Thermal defect resistance, therefore, allows us to compare measurements made \textit{\textit{in situ}} during ion irradiation at the irradiation temperature and those made ex-situ at room temperature following neutron irradiation. This approach is used in the upcoming discussion where thermal defect resistance of neutron irradiated 3C-SiC ($\alpha$ measured using LFA at room temperature before and after irradiation) and ion irradiated 3C-SiC ($\alpha$ measured using TGS at the irradiation temperature before and during irradiation) are compared.

\begin{figure}[h!]
    \centering
    \includegraphics[width=0.75\linewidth]{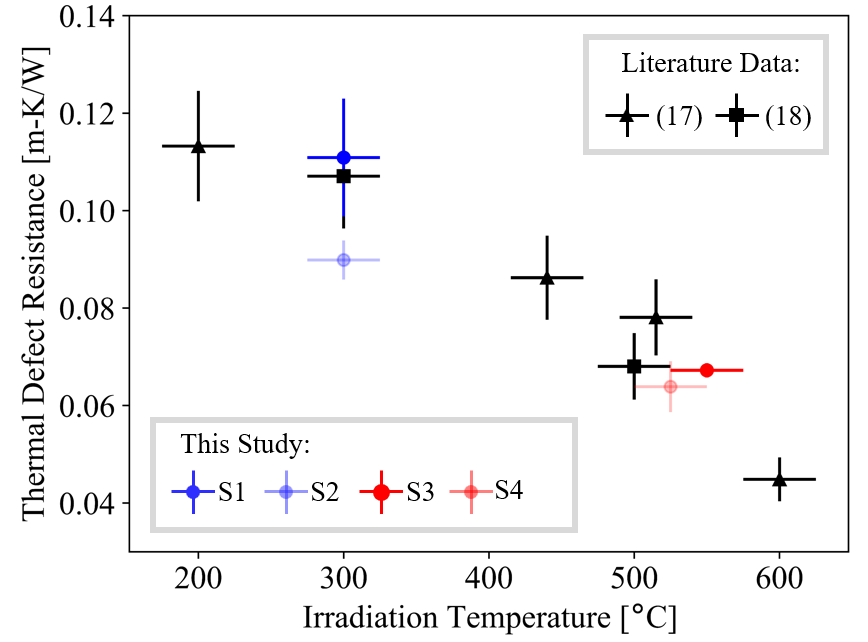}
    \caption{Thermal defect resistance of neutron and ion irradiated 3C-SiC as a function of irradiation temperature. The values for neutron irradiation were taken from \cite{snead_limits_2004, snead_handbook_2007}. Values for Liu (2025) were taken from \cite{liu_thermal_2025}}
    \label{fig:thermal_defect_resistance_vs_irradiation_temp}
\end{figure}

The thermal defect resistance of neutron and ion-irradiated CVD 3C-SiC as a function of irradiation is plotted in Figure \ref{fig:thermal_defect_resistance_vs_irradiation_temp} \cite{snead_limits_2004, snead_handbook_2007}. Given preceding discussions about the differences bulk and TGS measurements, it is expected that the thermal defect resistance evaluated using TGS and LFA will also not be equivalent. Knowing that thermal conductivity is expressed as $K  = \alpha C_p \rho$, Eq. \ref{eq:thermal_defect_resistance} can be modified to calculate the thermal defect resistance of ion irradiated samples (S1-S4) as measured by TGS:  

\begin{equation}\label{eq:thermal_defect_resistance_TGS}
    \frac{1}{K_{defect}} = \frac{S(\Lambda)}{C_p \rho} \left( \frac{1}{\alpha_\mathrm{irrad,TGS}} - \frac{1}{\alpha_{0,\mathrm{TGS}}} \right)
\end{equation}

\noindent where $\alpha_\mathrm{irrad,TGS}$ is the numerical average of the last ten thermal diffusivity measurements made using TGS during the 300 \degree C and 550 \degree C irradiations, while $\alpha_{0,\mathrm{TGS}}$ is the thermal diffusivity measured using TGS just before irradiation. $C_p$ is the heat capacity of high-purity SiC which was assumed to remain constant during irradiation and taken from \cite{snead_handbook_2007}. The mass density ($\rho$) was taken as 3.21 $\mathrm{g/cm^3}$ per the manufacturer's specification and was assumed to remain constant as a function of temperature \cite{snead_handbook_2007}. Although 3C-SiC is expected to undergo irradiation-induced lattice swelling of a few percent over the temperature range examined in this study, this effect is small compared to the large changes observed in thermal diffusivity \cite{snead_handbook_2007}. As a result, the effect of displacement damage on density has been ignored. $S({\Lambda})$ is a dimensionless `Suppression Factor' that was introduced in the previous section. Based on literature estimates for defect concentration in 3C-SiC after neutron and ion irradiation, the vacancy (or vacancy-type defect) concentration in CVD 3C-SiC samples irradiated in this study is expected to be greater than 1000 appm \cite{youngblood_effects_2004, liu_thermal_2025}. Accordingly, we apply $S(\Lambda)$ while noting that it introduces some error as described in Sections \ref{subsec:grating_spacing_thermal_diff} and \ref{sec:suppression_factor}. The final dose, effective thermal diffusivity, and suppression factor used are provided in Table \ref{tab:saturation_thermal_diff}.



\begin{table}[h]

\centering
\renewcommand{\arraystretch}{1.5} 

\begin{tabular}{
  >{\centering\arraybackslash}m{1cm}
  >{\centering\arraybackslash}m{1.5cm}
  >{\centering\arraybackslash}m{1.5cm}
  >{\centering\arraybackslash}m{1.5cm}
  >{\centering\arraybackslash}m{1.5cm}
  >{\centering\arraybackslash}m{1cm}
  >{\centering\arraybackslash}m{1.5cm}
  >{\centering\arraybackslash}m{1cm}
}
    \hline
  Sample & Irrad. Temp. [\degree C] & $\Lambda \: \mathrm{[\upmu m]}$ & $\alpha_\mathrm{TGS,0}$ $\mathrm{[mm^2/s]}$ & $c_{R,0}$ $\mathrm{[m/s]}$& Final Dose [dpa] & $\alpha_\mathrm{TGS,irr}$ $\mathrm{[mm^2/s]}$ & $S(\Lambda)$\\ 
  \hline
  S1 & 300 & 6.6318 & 27.57 $\pm$ 0.41 & 6854.29 $\pm$ 9.81 & 0.75 & 1.68 $\pm$ 0.18 & 0.65 \\
  S2 & 300 & 6.6512 & 32.47 $\pm$ 0.62 & 6978.14 $\pm$ 9.16 & 0.59 & 2.41 $\pm$ 0.11 & 0.77 \\
  S3 & 550 & 6.6604 & 16.51 $\pm$ 1.78 & 6740.25 $\pm$ 16.53 & 0.86 & 2.28 $\pm$ 0.05 & 0.65 \\
  S4 & 525 & 6.6234 & 20.38 $\pm$ 3.01 & 6862 $\pm$ 19.66 & 0.90 & 2.85 $\pm$ 0.23 & 0.76 \\
  \hline
\end{tabular}
\caption{Summary of irradiation conditions and pre- and post-irradiation properties}\label{tab:saturation_thermal_diff}
\end{table}

Figure \ref{fig:thermal_defect_resistance_vs_irradiation_temp} shows that there is good agreement between the thermal defect resistance of CVD 3C-SiC calculated using room temperature measurements of $\alpha_\mathrm{LFA}$ made before after neutron irradiation and measurements of $\alpha_\mathrm{TGS}$ made before and during and ion irradiation. The post-neutron irradiation room temperature thermal conductivities are taken near a dose of 1 dpa and a room temperature pre-irradiation thermal conductivity of $381 \: W/m/K$ has been assumed \cite{snead_limits_2004, snead_handbook_2007}. 
As noted in Section \ref{ref:thermomech_vs_dose}, $\alpha_\mathrm{TGS}$ data for S2 between 0.59 and 0.65 dpa were excluded from the analysis. As a result, the thermal defect resistance for S2 was evaluated at a lower dose than for S1, S3 and S4, possibly leading to a lower defect resistance than of S1, despite the two being evaluated at nearly the same temperature.

Although encouraging, the agreement between thermal defect resistance of ion and neutron irradiated 3C-SiC should be investigated further for two primary reasons. First, as the number of defects produced by the lower dose rate of neutron irradiation should be much smaller than that produced by the higher dose rate of ion irradiation. This discussion should be considered in the context of the dominant defect being vacancies and small vacancy clusters, since cavities and voids do not form in the given temperature and dose regime without the introduction of quantities of gasses such as helium \cite{hu_microstructural_2018, koyanagi_mechanical_2018}. Applying the Mansur relation to get an invariance in the number of defects lost to sinks at a fixed dose given a difference in dose rate (assuming $10^{-6}$ for neutron and $10^{-4}$ for ions), the required temperature shift is between 24 \degree C and 44 \degree C and between 41 \degree C and 83 \degree C  for the 300 \degree C and  500 \degree C neutron irradiations (the latter were provided as 480\degree C - 550\degree C in \cite{snead_handbook_2007}) respectively. Refer to Supplementary Material for more details on the temperature shift calculations. Despite two orders of magnitude higher dose rate, the required temperature shifts are small given the high migration energy of the Si and C vacancies \cite{gao_atomistic_2004}. It is therefore reasonable to see that the thermal defect resistance of the 300\degree C ion irradiated 3C-SiC is nearly identical to that of the neutron irradiation material at the same temperature, while the thermal defect resistance of the 550\degree C ion irradiated 3C-SiC matches that of the slightly lower 500 \degree C neutron irradiated material. For the thermal defect resistance of neutron irradiated SiC, error bars were taken from the standard deviation in thermal conductivity measurements provided by Katoh et al. while error bars in temperature were taken from Le Coq et al. \cite{katoh_stability_2011, le_coq_design_2018}. For the ion irradiated samples, error bars in the thermal defect resistance are taken from the standard deviation of the final ten $\alpha_\mathrm{TGS}$ measurements for each irradiation while the error bars for the irradiation temperature were determined based on a steady state thermal simulation in ANSYS Mechanical using Workbench 2025 R1 (refer to the Supplementary Material for more details).

Returning to the questions posed at the beginning of this section, it has been shown that $\alpha_\mathrm{TGS}$ can be used as a measure of changes in $\alpha$ of bulk CVD 3C-SiC following irradiations up to 1 dpa at 300 and 550\degree C. This result also indicates synergies between ion irradiation and fission neutron irradiation for CVD 3C-SiC in the temperature and dose ranges studied considered here. 

\section{Methods and Materials}\label{sec:Methods}

\subsection{Materials Preparation}

CVD 3C-SiC samples, measuring 10 mm x 10 mm x 1.5 mm, were purchased from PremaTech Advanced Ceramics. The manufacturer reports a purity of 99.9995\%, a nominal density of 3.21 $\mathrm{g/cm^3}$, and an average grain size of 5 \textmu m.
Samples for ion irradiation and TGS were mechanically polished down to 0.05 \textmu m using diamond suspensions of 9, 3 and 1 \textmu m, followed by a 0.05 \textmu m colloidal silica slurry. After polishing, samples were cleaned in acetone, isopropyl alcohol, and de-ionized water. The samples were then  baked at 850 $\degree$ C in a vacuum furnace. Subsequently coated with $\sim$ 36 nm of gold using a Polaron E5100 Sputter Coater or $\sim$ 20 nm of tungsten using the Kurt J. Lesker PRO Line PVD 75 Magnetron Sputtering Coater. Tungsten films were deposited using a 3-inch 99.5\% pure tungsten sputtering target from Kurt J. Lesker in an argon environment (3 mTorr, 22 sccm) using DC power of 40 W for 4.5 minutes. During deposition, samples remained at room temperature and were rotated at 8 RPM to promote uniform coating. 

\subsection{Transient Grating Spectroscopy}

Transient grating spectroscopy was performed on the Plasma Ion \textit{\textit{in situ}} Transient Grating Spectroscope (PI$^3$TGS) at MIT \cite{wylie_accelerating_2025, trachanas_study_2025}. A nominal thermal grating period of $\Lambda = 6.4$  \textmu m was chosen to balance two opposing constraints. First, given that ion damage is confined to a shallow near-surface layer (2 $\mathrm{\upmu m}$ in the present study, Figure \ref{fig:experimental_setup}), the thermal probe depth ($\Lambda/\pi$) should be as small as possible to maximize sensitivity to the irradiated region. Second, the SAW frequency ($\propto 1/\Lambda$) should be within the bandwidth of the Hamamatsu C5658 avalanche photodiode detectors used (50 kHz - 1 GHz) According to the detector manufacturer's specifications, signals with frequency equal to 1 GHz are attenuated by 3 dB while those approaching 2 GHz are attenuated by over 40 dB representing the sharp cutoff in frequencies greater than 1 GHz. Thus for materials with large SAW speeds, larger grating periods are needed.  
A 532 nm Q-switched laser was used as the `pump', having a repetition rate of 1 kHz, a pulse length of approximately 500 ps and a laser power between 5.52 and 7.36 mW. A continuous wave (CW) 577 nm laser was used as the `probe', chopped at a rate identical to that of the pump with a duty cycle of 20 \% using an optical chopper. The probe laser power was set between 0.12 and 0.14 mW. The superimposed diffracted and reflected probe beams (homodyne signal) were directed into avalanche photodiodes which convert the optical signal into an electrical signal. Each TGS measurement was averaged over 10,000 signals. The thermal diffusivity and SAW frequency were determined using a custom MATLAB program provided in the data repository with this publication \cite{short2025tgs}. For more details regarding the setup, refer to Wylie et al. \cite{wylie_accelerating_2025}. 

The setup was calibrated before each experiment using a single crystal tungsten with \{100\} surface orientation. 
Given that the SAW speed ($c_R$) of single crystal tungsten is known ($c_R = 2665.9 \: \mathrm{m/s}$) and the SAW frequency ($f$) can be extracted from the power spectral density of the signal, the experimental thermal grating period can be calculated as $\Lambda = c_R/f$ \cite{dennett_bridging_2016}. The numerical mean of five such measurements was then taken as the thermal grating period for the experiment. No changes were made to the optics upstream of the phase mask after this calibration.  

\subsection{Ion Irradiation}

\begin{figure}[h!]
    \centering
    \includegraphics[width=\linewidth]{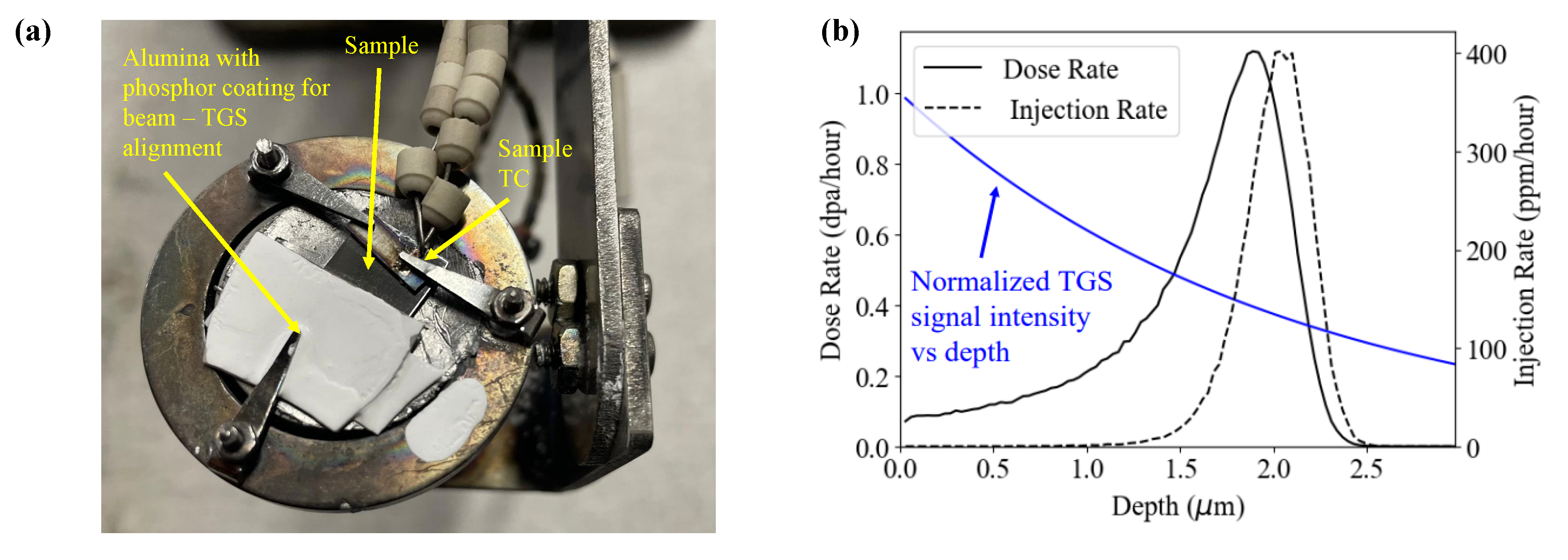}
    \caption{(a) Figure showing the sample setup before irradiation. (b) Dose and implantation profiles of 6.5 MeV $Si^{4+}$ ions into 3C-SiC calculated using SRIM}
    \label{fig:experimental_setup}
\end{figure}

Ion irradiation was performed at the Plasma Science and Fusion Center at the Massachusetts Institute of Technology using 6.5 MeV Si$^{4+}$ ions at 300\degree C (S1 and S2), 525\degree C (S4) and 550\degree C (S3), with sample temperature monitored using an on-sample thermocouple (TC) placed near the beam spot (Figure \ref{fig:experimental_setup} (a)). The ion beam was incident on the sample at approximately 45\degree while the axis of the TGS achromat lenses was normal to the sample. Note that in order to enable high-temperature operation, the sample TC was spot welded to a thin Titanium-Zirconium-Molybdenum alloy (TZM) piece which was then pressed onto the sample using clips. The TZM piece was ground down on the sample facing side using general purpose hand papers and washed in isopropyl alcohol prior to spot welding. Clips were used to hold the assembly in place. The sample was heated using a resistive button heater. A thin sheet of graphite was placed between the heater's top surface and the sample to minimize interfacial thermal resistance. Typical beam currents of 20 - 50 nA were measured using a Faraday cup placed near the sample, yielding an on-sample flux of approximately \mbox{$5\times10^{11}\,\mathrm{ions\,cm^{-2}\,s^{-1}}$}. The resulting dose rate and implantation profiles (Figure \ref{fig:experimental_setup} (b)) were calculated according to the method described by Short et al. and averaged over 8 independent simulations, each with 10,000 ions, performed using `Ion Distribution and Quick Calculation of Damage' in SRIM 2013 \cite{short_modeling_2016, ziegler_srim_2010}. The displacement threshold energy and mass density for SiC were set to 35 eV and 3.2 g/cm$^3$ \cite{katoh_microstructural_2006}. The normalized TGS signal intensity for thermal diffusivity ($e^{-z/ \Lambda}$) is also plotted in Figure \ref{fig:experimental_setup} (b). The dose [dpa] reported in the remainder of the paper is the numerical average of the dose over the thermal probe depth ($\Lambda/\pi$ $\approx$ 2 \textmu m), which represents the point at which the TGS signal intensity for thermal diffusivity has decayed to nearly 70 \% of the near-surface value \cite{dennett_capturing_2019}. At least ten TGS measurements were made before starting the irradiation. Each TGS signal was averaged over 10000 traces (1000 traces/second) and with a new TGS signal acquired every minute. 

\subsection{Variational solution to Boltzmann Transport Equation}

To describe heat conduction from a microscopic perspective in TGS, the Boltzmann transport equation (BTE) was used. Procedures from  prior work that experimentally validated variational solutions to the BTE in the TGS transmission and reflection geometries were followed \cite{huberman_unifying_2017, dennett_thermal_2018}. To obtain the inputs into the variational solutions, first-principles density functional theory calculations were performed. The DFT calculations were performed using PBEsol exchange-correlation functionals \cite{perdew_restoring_2008}. A 12 × 12 × 12 k-points mesh of Monkhorst-Pack with a 120 Ry kinetic energy cutoff and a convergence threshold of 
$10^{-12}$ Ry was chosen. Quantum ESPRESSO package \cite{QE-2009, QE-2017} was used to obtain second-order interatomic force constants, whereas thirdorder.py \cite{li_shengbte_2014} was used to obtain third-order interatomic force constants. A 6 × 6 × 6 supercell was used to calculate second-order interatomic force constants. A 3 × 3 × 3 supercell with fifth-nearest-neighbor interactions was used to calculate the third-order interatomic force constants. As in the case of silicon, the relaxation time approximation (RTA) was sufficient to describe the heat conduction in 3C-SiC \cite{lindsay_ab_2013, klemens_theory_1984}. Therefore, to account for different phonon scattering mechanisms, Matthiessen's rule was invoked:

\begin{equation}\label{eq:scattering_rates}
    \frac{1}{\tau_{total}} = \frac{1}{\tau_{p-p}}+\frac{1}{\tau_{vacancy}}+\frac{1}{\tau_{grains}}
\end{equation}

\noindent where $\tau_{p-p}$ is the phonon relaxation time due to three-phonon processes, $\tau_{vacancy}$ is the relaxation time due to vacancies scattering, and $\tau_{grains}$ is the relaxation time due to grain boundaries. $\tau_{p-p}$ and $\tau_{grains}$ are obtained using using the ShengBTE package \cite{li_shengbte_2014}. $\tau_{vacancy}$ is obtained following the analysis from \cite{gurunathan_analytical_2020}:

 \begin{equation}
    \frac{1}{\tau_{vacancy}} = \frac{1}{\tau_{mass}}+\epsilon\frac{1}{\tau_{radius}}
\end{equation}

\noindent where $\tau_{mass}$ is the relaxation time due to mass difference scattering as described by Tamura and $\tau_{radius}$ is the relaxation time due to atomic radius difference, analogous to the mass difference. Note, in all of the above, for a given relaxation time $\tau$, $1/\tau$ is the associated scattering rate. $\epsilon$ is a phenomenological fitting parameter that was set to 10. Using these scattering rates, thermal conductivity was converged on a 19 × 19 × 19 q-mesh with a Gaussian smearing parameter of 1.0 for the Kronecker delta approximation.

\subsection{Microstructural Characterization}

Electron backscatter diffraction (EBSD) was performed using a Zeiss Merlin High Resolution Scanning electron microscope (SEM) operated at an accelerating voltage of 20 kV, a probe current of 10 nA, and a working distance of 14.9 mm. Grain boundaries were identified by point-pairs between which the misorientation angle was greater than 2\degree. 

SEM imaging and FIB milling of sample S3 were performed using a FEI Helios 660. SEM micrographs were acquired at an accelerating voltage of 30 kV. FIB milling was done using 30 kV Ga ions at a current of 2.3 nA, followed by thinning at 0.79 nA (1\degree tilt) and 80 pA (2\degree  tilt). Final milling was done using 5 kV Ga ions at 3\degree undertilt and a current of 41 pA over a box size of 11.64 \textmu m $\times$ 1.38 \textmu m to minimize FIB damage. Transmission electron microscopy (TEM) and energy dispersive X-Ray spectroscopy (EDS) were performed using a JEOL ARM 200F STEM operated at an accelerating voltage of 200 kV. EDS linescans were captured with the 3C probe, using spatial resolution of 5 points/nm, energy resolution of 10 eV/channel, 2048 frames, and a dwell time of 100 ms. 

\subsection{Laser Flash Analysis}

Bulk thermal diffusivity measurements were obtained using a Netzsch \\ LFA467 HT. Prior to measurements, polished SiC samples were coated with a graphite spray to limit surface reflectivity and improve signal-to-noise ratio. For the temperature study, LFA measurements were made in steps of 25 \degree C between room temperature and 800\degree C, with each datapoint representing the numerical average of 3 measurements. 

\section{Supplementary Material}

Refer to this link for the \href{https://www.dropbox.com/scl/fi/r2ufx4nnj8zw3p6z14qrq/Supplementary-Material.docx?rlkey=05p3xgke38ui4t30xu2hq18hp&st=vlg7gf47&dl=0}{Supplementary Material}.

\section{Data availability statement}

Experimental and computational data, including all TGS traces and micrographs, that support findings of this study are freely available at the online repository, . Scripts used to run the computational DFT+BTE models will be provided upon reasonable request to the corresponding author. 

\section{Funding Source}

This work was funded by Eni S.p.A through the MIT Laboratory for Innovation in Fusion Technologies.

\section{Credit authorship contribution statement}

\textbf{Keshav Vasudeva} Conceptualization, Methodology, Formal Analysis, Investigation, Writing - Original Draft, Writing - Review \& Editing, Visualization.  \textbf{Samuel Huberman} Methodology, Formal Analysis, Investigation, Writing - Original Draft, Writing - Review \& Editing. \textbf{Angus P. C. Wylie} Methodology, Investigation, Writing - Review \& Editing. \textbf{Angus P. C. Wylie} Methodology, Investigation, Writing - Review \& Editing. \textbf{Joey Demiane} Methodology, Writing - Review \& Editing. \textbf{Jamal A. Haibeh} Methodology, Writing - Review \& Editing. \textbf{Elena Botica-Artalejo} Methodology, Writing - Review \& Editing. \textbf{Kevin B. Woller} Methodology, Writing - Review \& Editing. \textbf{Michael P. Short} Conceptualization, Methodology, Supervision, Writing - Review \& Editing. \textbf{Sara E. Ferry} Conceptualization, Methodology, Supervision, Funding acquisition, Writing - Review \& Editing. 

\section{Declaration of generative AI and AI-assisted technologies in the writing process}

During the initial preparation of this work, a subset of authors used ChatGPT-5 model for English correction and expression. ChapGPT-5 was also used to debug the scripts used to post-process and plot the raw data. After using this tool/service, all authors reviewed and edited the content as needed, and take full responsibility for the content of the publication.  

\section{Declaration of Competing Interest} 

The authors declare that they have no known competing financial interests or personal relationships that could have appeared to influence the work reported in this paper.

\section {Acknowledgments}

We would like to thank Charlie Settens at MIT Nano for help with EBSD. This work was carried out in part through the use of MIT.nano's facilities. This work was performed in part at the Harvard University Center for Nanoscale Systems (CNS); a member of the National Nanotechnology Coordinated Infrastructure Network (NNCI), which is supported by the National Science Foundation under NSF award no. ECCS-2025158. We would also like to thank Axel Beamwright for big beautiful beams.

\printbibliography


\end{document}